\documentclass[prd,showpacs,preprintnumbers,amsmath,amssymb]{revtex4}

\usepackage{graphicx}
\usepackage{dcolumn}
\usepackage{bm}
\usepackage{latexsym}
\usepackage{amsfonts}
\usepackage{amssymb}

\newcommand{\R }{ \overset{(4)}{R}{}^{\mu}_{\ \nu} }
\newcommand{\Rs }{ \mathop{R}^{(4)}  }

\begin{document}

\preprint{KUCP0226}

\title{Low Energy Effective Action for Dilatonic Braneworld\\ 
--A Formalism for Inflationary Braneworld--}

\author{Sugumi Kanno}
\email{sugumi@tap.scphys.kyoto-u.ac.jp}

\author{Jiro Soda}
\email{jiro@tap.scphys.kyoto-u.ac.jp}
\affiliation{
 Department of Physics, Kyoto University, Kyoto 606-8501, Japan
}%

\date{\today}

\begin{abstract}
 We derive the low energy effective action for the dilatonic braneworld.
 In the case of the single-brane model, we find 
 the effective theory is described 
 by the  Einstein-scalar theory coupled to 
 the dark radiation. Remarkably, the dark radiation is not conserved 
 in general due to a coupling to the bulk scalar field.
   The effective action incorporating  Kaluza-Klein (KK)
 corrections is  obtained and the role of the AdS/CFT correspondence 
 in the dilatonic braneworld is revealed.  
 In particular, it is shown that CFT matter would not be 
  confined to the braneworld in the presence of the bulk scalar field. 
 The relation between our analysis and the geometrical projection method is
 also clarified.  In the case of the two-brane model, 
  the effective theory reduces to a scalar-tensor
 theory with a non-trivial coupling between the radion and the bulk scalar
 field.    
\end{abstract}

\maketitle

\section{Introduction}

Theoretically, it is usual to assume the existence of extra dimensions.
 In fact, the superstring theory requires  ten spacetime dimensions to keep
 consistency. Empirically, however, we know our universe is  a
  four-dimensional spacetime. To reconcile these apparently contradicted
 views, we need a concept of the compactification. Inspired by the recent 
 developments of the superstring theory~\cite{braneworld}, 
 Randall and Sundrum (RS) have proposed
 a novel compactification mechanism in the context of the braneworld 
 scenario~\cite{RS1}. Interestingly, in this mechanism, gravity  can be 
 localized on the brane due to the warped geometry, 
 which allows even the non-compact extra dimensions. Because of this attractive
 feature, there are many works on cosmology and black hole 
 in a braneworld~\cite{others1,others2}. 
 
   In the superstring theory,  scalar fields  are 
 ubiquitous. Indeed, a dilaton and moduli exist in the bulk 
 generically, because they arise as   modes associated with a 
 closed string. Moreover, when the supersymmetry is spontaneously broken,
 they may have a non-trivial potential. Hence, it is natural to incorporate
  a bulk scalar field with non-trivial potential into the RS model, 
  which is often called dilatonic braneworld. In addition to this theoretical
  motivation, there is a phenomenological interest in this model. 
  Namely, the bulk scalar field can drive the inflation on the brane
  although the bulk spacetime never inflate~\cite{bulk,sago}. 
  This kind of inflationary 
  scenario  certainly deserves further investigations. 
   
  In any case, it is nice if a purely 4-dimensional description of the
  braneworld exists. In the case of the non-dilatonic braneworld, there are
  two independent approaches to obtaining the 4-dimensional effective theory.
  One approach is to use a geometrical projection method~\cite{ShiMaSa} 
  which is proved to
  be useful for understanding cosmological perturbations~\cite{soda}.
  However, it is not a closed system of equations. In fact, the projected
  Weyl tensor $E_{\mu\nu}= C_{y\mu y\nu}|_{y=0}$ 
  which represents the effect of the bulk geometry
  can not be determined within this  theory.
  In other words, there exists no action for this projected equations of motion.
  The other approach is  based on the AdS and conformal field theory (CFT) 
  correspondence~\cite{ads}. 
  There exists an action for this effective theory, however, it is valid only
  at low energy.  This approach also gives rise  to an interesting 
  consequence~\cite{tanaka2}. Because both approaches are complementary and
  useful, their mutual relations should be understood. 
  This was done by us through the explicit construction of the low energy
  effective action~\cite{kanno1}. 
  It would be beneficial if we could generalize these results to the dilatonic 
  braneworld.
    
  Fortunately, the extension of the geometrical projection method to the 
  dilatonic case has been already known~\cite{maeda} (see also \cite{uzawa}).
  The 4-dimensional effective Einstein equation reads 
\begin{eqnarray}
    G_{\mu\nu}=
	{2\kappa^2\over 3}\partial_\mu\varphi\partial_\nu\varphi
	-{5\kappa^2\over 12}g_{\mu\nu} 
	\partial^\alpha\varphi\partial_\alpha\varphi
	-\Lambda(\varphi)g_{\mu\nu} 
	+{\kappa^4\over 6}\sigma(\varphi)T_{\mu\nu}
	+\kappa^4\pi_{\mu\nu}-E_{\mu\nu}  \ , 
	\label{mw:metric}
\end{eqnarray}
where
\begin{eqnarray}  
\Lambda(\varphi)={\kappa^2\over2}\left[U(\varphi)
	+{\kappa^2 \over 6} \sigma^2(\varphi) 
      	-{1 \over 8}\left(\frac{d\sigma(\varphi)}{d\varphi}\right)^2 
      	\right]       \ .
\end{eqnarray}
Here, $\pi_{\mu\nu}$, $U$ and $\sigma$ are the quadratic of the energy 
momentum tensor, the bulk potential and the brane tension, respectively. 
 Curiously, 
the field $\varphi$ has non-conventional kinetic terms, hence it can not be 
derived from an action. This prevent us from interpreting 
$\Lambda (\varphi) $ as the potential function. 
Therefore, the geometrical approach may have disadvantages in applying it to the
 physical problems, although it has a clear geometrical meaning. 
 This urges us to find the AdS/CFT interpretation 
 of the dilatonic braneworld. Since it is an action approach, the enigma
 of kinetic terms for the field $\varphi$ would be resolved, which lead us
 to more profound understanding of the dilatonic braneworld. 

  Thus, the purpose of this paper is to construct the low energy effective 
  action and make AdS/CFT interpretation of the dilatonic braneworld.
  In the course of analysis, we clarify the relation between the geometrical
  approach and the AdS/CFT approach. Apart from these conceptual developments,
  we present the action of two-brane system as a concrete result. 
   
 The organization of this paper is as follows.
 In sec.2, we present the action for the dilatonic braneworld
  and derive  basic  equations. 
  In sec.3, the low energy approximation scheme is explained briefly.
 The non-linear low energy effective actions for both  
 the single-brane model and
 the two-brane model are derived. The relation to the geometric projection 
 method is clarified. 
 In sec.4,  the effective action with KK corrections are derived
  and its implications  are discussed. In particular, 
  the role of AdS/CFT correspondence is revealed and the relation to
  the geometrical projection method is further investigated. 
 The final section is devoted to the conclusion. 
 In the appendix A,  useful formula for the calculation are displayed. 
 In the Appendix B, the details of the calculation at  second order
 can be found.  

\section{Model and Basic Equations}

Inflation in the braneworld can be driven by a scalar field 
either on the brane or in the bulk.
We derive the effective equations of motion which are useful for both 
models. 
In this section, we begin with the single-brane system. Since we 
know the effective 4-dimensional equations hold irrespective of the existence
 of other branes~\cite{kanno2},
  the analysis of the single-brane system is sufficient
 to derive the effective action for the two-brane system 
 as we see in the next section. 

We  consider a $Z_2$ symmetric 5-dimensional spacetime with a  
 brane at the fixed point 
 and a  bulk scalar field $\varphi$ 
coupled to the brane tension $\sigma(\varphi)$
 but not to the matter ${\cal L}_{\rm matter}$ on the brane. 
 The correponding action  is given by   
\begin{eqnarray}
S &=& {1\over 2\kappa^2}\int d^5 x \sqrt{-g}~{\cal R}
	-\int d^5 x  \sqrt{-g} \left[~ 
	{1\over 2} g^{AB} \partial_A\varphi 
	\partial_B\varphi
	+ U(\varphi) ~\right]\nonumber\\
&&	-\int d^4 x \sqrt{-h}~
	\sigma(\varphi)
	+\int d^4 x \sqrt{-h}~ 
	{\cal L}_{\rm matter} \ ,
\label{eq:action}
\end{eqnarray}
where $\kappa^2$, ${\cal R}$ and $h_{\mu\nu}$ are the gravitational constant, 
the scalar curvature in 5-dimensions constructed from the metric $g_{AB}$
 and the induced metric on the brane, respectively.  
 We assume the potential $U(\varphi)$ for the bulk scalar field takes the form
\begin{eqnarray}
U(\varphi)&=&-\frac{6}{\kappa^2\ell^2}+V(\varphi)\ ,
\label{potential}
\end{eqnarray}
where the first term is regarded as a 5-dimensional cosmological constant
and the second term is an arbitrary potential function.
The brane tension is also assumed to take the form
\begin{eqnarray}
\sigma(\varphi)&=&\sigma_0+\tilde\sigma(\varphi)\ .
\end{eqnarray}
The constant part of the brane tension, $\sigma_0$ is tuned so that
the effective cosmological constant on the brane vanishes. 
The above setup realizes a flat braneworld after inflation ends and 
the field $\varphi$ reaches the minimum of its potential. 

We adopt the Gaussian normal coordinate system to describe 
the geometry of the brane model;
\begin{equation}
ds^2 = dy^2 + g_{\mu\nu} (y,x^{\mu} ) dx^{\mu} dx^{\nu}  \ ,
\label{eq:5-dim-metric}
\end{equation}
where the brane is assumed to be located at $y=0$. 
 Let us decompose the extrinsic curvature into the traceless part 
 $\Sigma_{\mu\nu}$ and the trace part $K$ as
\begin{equation}
K_{\mu\nu}=-\frac{1}{2}g_{\mu\nu,y}
	=\Sigma_{\mu\nu}
	+{1\over 4} g_{\mu\nu} K \  .
\label{eq:decompose} 
\end{equation}
Then, we can obtain the basic equations off the brane 
using these variables.
First, the Hamiltonian constraint equation leads to
\begin{eqnarray}       		
{3\over 4} K^2-\Sigma^{\alpha}{}_{\beta}\Sigma^{\beta}{}_{\alpha}
	&=&\Rs -\kappa^2 \nabla^\alpha\varphi
	\nabla_\alpha\varphi 
	+\kappa^2(\partial_y\varphi )^2
	-2\kappa^2U(\varphi ) \ ,  
    	\label{eq:hamiltonian} 
\end{eqnarray}
where $\R$ is the curvature on the brane and $\nabla_\mu $ denotes the
covariant derivative with respect to the metric $g_{\mu\nu}$. Momentum 
constraint equation becomes
\begin{eqnarray}       		
\nabla_\lambda \Sigma_{\mu}{}^{\lambda}  
	-{3\over 4} \nabla_\mu K =-\kappa^2
	\partial_y\varphi\partial_\mu\varphi\ .
	\label{eq:momentum}  
\end{eqnarray}
Evolution equation in the direction of $y$ is given by
\begin{eqnarray}
\Sigma^{\mu}{}_{\nu,y}-K\Sigma^{\mu}{}_{\nu} 
	=-\left[ \R-\kappa^2 \nabla^\mu\varphi
	\nabla_\nu\varphi 
	\right]_{\rm traceless}      \ .     
	\label{eq:evolution} 
\end{eqnarray}
Finally,  the equation of motion for the scalar field reads
\begin{eqnarray}
\partial^2_y\varphi -K\partial_y\varphi 
	+ \nabla^{\alpha}\nabla_{\alpha}\varphi -U'(\varphi)=0  \ ,
	\label{eq:scalar}	
\end{eqnarray}
where the prime denotes derivative with respect to the scalar field $\varphi$.

 As we have the singular source at the brane position, we must consider 
 the junction conditions. 
 Assuming a $Z_2$ symmetry of spacetime, we obtain the junction 
conditions for the metric and the scalar field 
\begin{eqnarray}
\left[ \Sigma^{\mu}{}_{\nu}-{3\over 4} \delta^\mu_\nu K \right] \Bigg|_{y=0}
	&=& -{\kappa^2 \over 2}\sigma(\varphi)\delta^\mu_\nu 
    	+\frac{\kappa^2}{2}T^{\mu}{}_{\nu}  \ , 
    	\label{JC:metric} \\
	\Bigl[\partial_y\varphi\Bigr]\bigg|_{y=0} &=& 
	{ 1\over 2} \sigma'(\varphi) \ ,
	\label{JC:scalar}         
\end{eqnarray}
where $T^{\mu}{}_{\nu}$ is the energy-momentum tensor for the matter
fields  on the brane.

\section{Low Energy Effective Action}

In order to derive the effective action,
 we take the following strategy.
  First, we solve the bulk equations and obtain the 
 bulk fields as the functional of the induced metric and the scalar field on 
 the brane. After that, we will impose the junction conditions on the solutions
 which can be regarded as conditions on the induced fields. 
 Thus, we obtain the effective equations of motion from which one can read off
 the effective 4-dimensional action for the dilatonic braneworld. 
 Needless to say, it is intractable to solve the bulk equations of motion
  in general. However, most interesting phenomena occur at low energy
 in the sense that the additional energy due to the bulk scalar field is small, 
 $\kappa^2\ell^2V(\varphi)\ll 1$, and the curvature on the brane $R$ 
 is also small, $R\ell^2\ll 1$.  It should be stressed that the low energy
 does not necessarily implies weak gravity on the brane. 
 Under these circumstances, we can use a
 gradient expansion scheme to solve the bulk equations of motion.  
  Then, following the above procedure, we can derive the effective low 
  energy action with corrections coming from Kaluza-Klein modes. 
  Let us explain this more concretely (see \cite{kanno1} for detailed 
  calculations and discussions). 

 At zeroth order,  we take the brane tension $\sigma(\varphi)$ to be constant 
 $\sigma_0$ and  ignore matters on the brane.
  Then, from the junction condition (\ref{JC:metric}), we have 
\begin{eqnarray} 
\left[ \overset{(0)}{\Sigma}{}^{\mu}{}_{\nu}
 -{3\over 4} \delta^\mu_\nu \overset{(0)}{K} \right] \Bigg|_{y=0}
	&=& -{\kappa^2 \over 2}\sigma_0 \delta^\mu_\nu  \ .
	\label{JC:metric0}
\end{eqnarray}
 As the right hand side of (\ref{JC:metric0}) contains no traceless part,
 we get 
\begin{eqnarray}
\overset{(0)}{\Sigma}{}^\mu{}_\nu  =0 \ .
\end{eqnarray}
We also take the potential for the bulk scalar field 
$U(\varphi)$ to be $-6/(\kappa^2\ell^2)$. 
We discard the terms with  4-dimensional derivatives since one can neglect
 the long wavelength variation in the direction of $x^\mu$ at low energies.  
 Thus, the equations to be solved are given by
\begin{eqnarray}       		
  && {3\over 4} \overset{(0)}{K}{}^2
	=  	  \kappa^2(\partial_y \overset{(0)}{\varphi} )^2
	  + {12 \over \ell^2 } \ ,  \\
    	\label{eq:hamiltonian0} 
  && \partial^2_y \overset{(0)}{\varphi} 
        - \overset{(0)}{K} \partial_y  \overset{(0)}{\varphi} =0  \ .
	\label{eq:scalar0}	
\end{eqnarray}
The  junction condition (\ref{JC:scalar}) at this order
\begin{eqnarray}
  \Bigl[\partial_y \overset{(0)}{\varphi} \Bigr]\bigg|_{y=0} = 0
  \label{JC:scalar0}
\end{eqnarray}
tells us that the solution of Eq.(\ref{eq:scalar0}) must be
\begin{eqnarray}
\stackrel{(0)}{\varphi}=\eta~(x^\mu) \ ,
\end{eqnarray}
where $\eta (x^\mu )$ is an arbitrary constant of integration.
 Now, the solution of  Eq.(\ref{eq:hamiltonian0}) yields
\begin{eqnarray}
\overset{(0)}{K}=\frac{4}{\ell} \ .
\end{eqnarray}
Other Eqs. (\ref{eq:momentum}) and (\ref{eq:evolution})
 are trivially satisfied at zeroth order. 
Using the definition
$\overset{(0)}{K}{}_{\mu\nu}= -\overset{(0)}{g}_{\mu\nu,y}/2$,  
we have the lowest order metric 
\begin{equation}
\overset{(0)}{g}{}_{\mu\nu} (y,x^\mu ) 
  = b^2(y)~h_{\mu\nu} (x^\mu ), \qquad 
b(y)\equiv e^{- y/ \ell} \ ,
\label{formal:metric}
\end{equation}
where the induced metric on the brane, $ h_{\mu\nu}\equiv g_{\mu\nu} 
(y=0 ,x^\mu) $, arises as a constant of  integration. 
The junction condition for the induced metric (\ref{JC:metric0}) merely 
implies well known relation $\kappa^2\sigma_0=6/\ell$
and that for the scalar field (\ref{JC:scalar0}) is trivially satisfied.
At this leading order analysis, we can not determine the constants of 
integration $h_{\mu\nu} (x^\mu)$ and $\eta (x^\mu)$ which are constant
 as far as the short length scala $\ell$ variations are concerned, but
 are allowed to vary over the long wavelength scale. These constants
 should be  constrained by the next order analysis. 

 Now, we take into account the effect of both the bulk scalar field 
and the matter on the brane perturbatively.
Our iteration scheme 
is to write the metric $g_{\mu\nu}$ and the scalar field $\varphi$ 
as a sum of local tensors built out of the  induced metric and the induced 
 scalar field on the brane,  in the order of expansion parameters, that is, 
$O((R \ell^2 )^{n})$ and $O(\kappa^2\ell^2V(\varphi))^n$, 
$n=0,1,2,\cdots$~\cite{tomita}. 
 Then, we expand the metric and the scalar field  as 
\begin{eqnarray}
&& g_{\mu\nu} (y,x^\mu ) =
	b^2 (y) \left[ h_{\mu\nu} (x^\mu) 
  	+ \overset{(1)}{g}_{\mu\nu} (y,x^\mu)
      	+ \overset{(2)}{g}_{\mu\nu} (y, x^\mu ) + \cdots  \right]  \ , 
      	\label{expansion:metric} \nonumber\\
&& \varphi(y,x^\mu) = \eta (x^\mu) 
	+ \overset{(1)}{\varphi} (y, x^\mu) +\overset{(2)}{\varphi} (y, x^\mu)
	+ \cdots  \ .
\end{eqnarray}
Here, we put the boundary conditions
\begin{eqnarray}
&& \overset{(i)}{g}_{\mu\nu} (y=0 ,x^\mu ) =  0  
	\ , \quad i=1,2,3,... 
   	\label{BC:metric}\\
&& \overset{(i)}{\varphi } (y=0 ,x^\mu) =0 \ ,
	\label{BC:scalar}
\end{eqnarray}
so that we can interpret $h_{\mu\nu}$ and $\eta$ as induced quantities. 
Extrinsic curvatures can be also expanded as 
\begin{eqnarray}
K &=& \frac{4}{\ell} 
	+\overset{(1)}{K}
	+\overset{(2)}{K}
	+\cdots  \ , \\
\Sigma^\mu_{\ \nu} &=& \qquad
	\overset{(1)}{\Sigma}{}^{\mu}_{\ \nu}
        +\overset{(2)}{\Sigma}{}^{\mu}_{\ \nu}
        +\cdots \ .
        \label{expansion:sigma}
\end{eqnarray}
Below, we discuss the single-brane model and the
two-brane model, separately.

\subsection{Single-brane Model}

The 5-dimensional geometry is deformed by the additional bulk scalar field 
and matter field on the brane. We show that junction conditions at first order 
lead to the effective equations on the brane and determine this deformation.
 
 At the first order, the Hamitonian constraint equation 
 (\ref{eq:hamiltonian})  becomes 
\begin{eqnarray}
\overset{(1)}{K}=
	{\ell\over 6}\left[ \overset{(4)}{R}
	-\kappa^2\nabla^\alpha\varphi\nabla _\alpha\varphi\right]^{(1)}
	-\frac{\ell}{3}\kappa^2 V(\eta) \ .
	\label{2-1:hamiltonian}
\end{eqnarray}
 Using the formula such as 
$\overset{(4)}{R} (\overset{(0)}{g}_{\mu\nu})=R(h_{\mu\nu})/b^2$, 
we obtain the solution
\begin{eqnarray}
\overset{(1)}{K}=\frac{\ell}{6b^2}\left(
	R (h) -\kappa^2\eta^{|\alpha}\eta_{|\alpha}\right)
	-\frac{\ell}{3}\kappa^2V(\eta) \ ,
	\label{1:hamiltonian} 
\end{eqnarray}
where $R(h)$ is the scalar curvature of $h_{\mu\nu}$
 and $|$ denotes the covariant derivative with respect to  
$h_{\mu\nu}$. 
 The evolution equation (\ref{eq:evolution}) at this order reads
\begin{eqnarray}
\overset{(1)}{\Sigma}{}^\mu{}_{\nu,y} 
	-{4\over\ell}\overset{(1)}{\Sigma}{}^\mu{}_\nu 
	= -\left[\overset{(4)}{R}{}^{\mu}{}_{\nu} 
	-\kappa^2\nabla^\mu\varphi\nabla _\nu\varphi\right]_
	{\rm traceless}^{(1)} \ .
 	\label{2-1:evolution}
\end{eqnarray}
Substituting the results at zeroth order solutions into 
Eq.~(\ref{2-1:evolution}), we obtain
\begin{eqnarray}
\overset{(1)}{\Sigma}{}^{\mu}{}_{\nu}=
	\frac{\ell}{2b^2}\left[
	R^{\mu}{}_\nu (h) -\kappa^2\eta^{|\mu}\eta_{|\nu}\right]_{\rm traceless}
	+\frac{\chi^{\mu}{}_\nu}{b^4} \ ,
	\label{1:evolution}
\end{eqnarray}
where $R^{\mu}{}_\nu (h)$ denotes the Ricci tensor of $h_{\mu\nu}$ and 
$\chi^{\mu}{}_\nu$ is a constant of integration which 
satisfies the constraint $\chi^{\mu}{}_{\mu}=0$. 
 Hereafter, we omit the argument of the curvature for simlicity. 
 Integrating the scalar field equation (\ref{eq:scalar}) at first order
\begin{eqnarray}
\partial_y^2\overset{(1)}{\varphi} 
	-{4\over \ell}\partial_y \overset{(1)}{\varphi}
	= -\left[\nabla^{\alpha}\nabla_{\alpha}\varphi\right]^{(1)}
 	 \ ,
\end{eqnarray}
we have
\begin{eqnarray}
\partial_y\overset{(1)}{\varphi}&=&\frac{\ell}{2b^2}\Box\eta
	-\frac{\ell}{4}V'(\eta)
	+\frac{C}{b^4} \ ,
	\label{1:scalar}
\end{eqnarray}
where $C$ is also a constant of integration. 
At first order in this iteration scheme,
we get two kinds of constants of integration, $\chi^{\mu}{}_{\nu}$ and $C$.

Given the brane tension $\tilde\sigma (\eta )$   
and the matter fields $T_{\mu\nu}$ on the brane,  the junction 
condition (\ref{JC:metric}) becomes
\begin{eqnarray}
\left[\overset{(1)}{\Sigma}{}^{\mu}{}_{\nu}
	-\frac{3}{4}\delta^\mu_\nu
	\overset{(1)}{K}\right] \Bigg|_{y=0} 
	=-\frac{\kappa^2}{2}\delta^{\mu}_{\nu}\tilde\sigma
	+\frac{\kappa^2}{2}T^{\mu}{}_{\nu} \ .
	\label{1:JC-m}
\end{eqnarray}
Substituting the solutions (\ref{1:hamiltonian}) and 
(\ref{1:evolution}) into the junction condition 
(\ref{1:JC-m}), we obtain the effective equation 
\begin{eqnarray}
&&G^{\mu}{}_{\nu}=\frac{\kappa^2}{\ell}T^{\mu}{}_{\nu}
	+\kappa^2\left[\eta^{|\mu}\eta_{|\nu}
	-\frac{1}{2}\delta^{\mu}_{\nu}\eta^{|\alpha}\eta_{|\alpha}
	-\delta^{\mu}_{\nu}V_{\rm eff}\right]
	-\frac{2}{\ell}\chi^{\mu}{}_{\nu} \ ,
	\label{1:einstein}
\end{eqnarray}
where we have defined an effective potential 
\begin{eqnarray}
V_{\rm eff}=\frac{1}{\ell}\tilde\sigma
	+\frac{1}{2}V  \ .
	\label{1:effpt}
\end{eqnarray}
We note that $\chi^{\mu}{}_{\nu}$ corresponds to ``dark radiation" 
in the case of homogeneous cosmology,
so we also refer to $\chi^{\mu}{}_{\nu}$ as ``dark radiation". 
This is the standard 4-dimensional Einstein-scalar 
equations with dark radiation. 
Strictly speaking, the induced scalar field $\eta$ should be normalized
 so as to have standard dimension $\eta^{(4)}=\eta/\sqrt{\ell}$. 
 As it does not cause any confusion, however, we leave it in the original form 
 in this paper. 

 At this order, the junction condition (\ref{JC:scalar}) yields
\begin{eqnarray}
\biggl[\partial_y\overset{(1)}{\varphi}\biggr] \Bigg|_{y=0}
	=\frac{1}{2}~\tilde\sigma' \ .
	\label{1:JC-s}
\end{eqnarray}
After substituting the solution (\ref{1:scalar}) into the junction condition 
 (\ref{1:JC-s}) and using the effective potential (\ref{1:effpt}),  
we obtain the Klein-Gordon equation 
\begin{eqnarray}
\Box\eta-V'_{\rm eff}=-\frac{2}{\ell} C \ ,
\label{1:KG}
\end{eqnarray}
where we refer to $C$ as ``dark source" originated from the bulk
scalar field.
The momentum constraint (\ref{eq:momentum}) gives constraint on
dark radiation and dark source 
\begin{eqnarray}
\chi^{\nu}{}_{\mu|\nu}=-\kappa^2~C~\eta_{|\mu} \ .
\label{1:momentum}
\end{eqnarray}
Unlike the non-dilatonic theory, 
 the dark radiation is not conserved due to the existence 
of the bulk scalar field. This result is consistent with 
the previous analysis~\cite{tanaka}.
The Bianchi identity applied to  Eq.~(\ref{1:einstein})
gives the conservation of the energy-momentum tensor for the matter fields, 
 $T^{\mu}{}_{\nu|\mu}=0$ provided the relation (\ref{1:momentum}).

 Now, the action can be read off from the Einstein equation, 
\begin{eqnarray}
S={\ell\over 2\kappa^2}\int d^4x\sqrt{-h}\left[ R
	-\kappa^2\eta^{|\alpha}\eta_{|\alpha}
	-2\kappa^2V_{\rm eff}\right]
	+\int d^4 x \sqrt{-h}~ 
	{\cal L}_{\rm matter}
 	+S_\chi \ , 
\end{eqnarray}
where  $S_\chi$ is defined by
\begin{eqnarray}
\frac{1}{\sqrt{-h}}{\delta S_\chi \over \delta h^{\mu\nu}}
	=\frac{1}{\kappa^2}~\chi_{\mu\nu} 
\end{eqnarray}
and
\begin{eqnarray}
\frac{1}{\sqrt{-h}}{\delta S_\chi \over \delta \eta}=2~C \ .
\end{eqnarray}
 These tell us  that the dark radiation couples with the scalar field
 through $C$. The relation (\ref{1:momentum}) 
guarantees the diffeomorphism invariance of $S_\chi$ (see Appendix A).
 
 To determine $\chi^{\mu}{}_{\nu}$ and $C$, we need the boundary condition
 at the AdS horizon.   One natural choice is to 
  impose the regularity at the AdS horizon.
   In the non-dilatonic case, the dark radiation $\chi^{\mu}{}_{\nu}$
  corresponds to the black hole    singularity in the bulk. 
  In the dilatonic case, Eq.(\ref{1:momentum}) implies
 $C$ induces $\chi^{\mu}{}_{\nu}$.  Hence, 
  to keep the regularity at the AdS horizon, we must impose  
 $\chi^{\mu}{}_{\nu} = C=0$ at this order.  Thus, $S_\chi =0$.
  When $S_\chi =0$, the inflation occurs on  the brane under the
slow-roll condition, $V''\ll H^2$. Moreover, it is obvious the standard
 primordial curvature fluctuations with the almost scale invariant spectrum
 is generated. Therefore, at least at this order, the braneworld inflation
 leads to the  prediction consistent with WMAP data~\cite{WMAP}.  

Using the solutions (\ref{1:hamiltonian}) and (\ref{1:evolution}), 
first order metric is given by
\begin{eqnarray}
\stackrel{(1)}{g_{\mu\nu}}&=&-\frac{\ell^2}{2}\left(
	\frac{1}{b^2}-1\right)
	\left[
	R_{\mu\nu}-\frac{1}{6}h_{\mu\nu}R-\kappa^2
	\left(\eta_{|\mu}\eta_{|\nu}
	-\frac{1}{6}h_{\mu\nu}\eta^{|\alpha}\eta_{|\alpha}\right)
	\right] \nonumber\\
&&	-\frac{\ell}{2}\left(\frac{1}{b^4}-1\right)\chi_{\mu\nu}(x)
	-\frac{\ell^2}{6}\log b~h_{\mu\nu}\kappa^2V(\eta) \ ,
\end{eqnarray}
where we imposed the boundary condition (\ref{BC:metric}). 
This tells us how we construct the bulk geometry using the 4-dimensional 
 data. 

Let us now compare our results with the geometrical approach by 
Maeda and Wands where the projected Weyl tensor 
$E_{\mu\nu}=C_{y\mu y\nu}|_{y=0}$ and $\Phi_2=\partial^2_y\varphi|_{y=0}$  
give us information about the bulk. At first order, 
their equations take the form
\begin{eqnarray}
G_{\mu\nu}&=&
	\frac{2\kappa^2}{3}\eta_{|\mu}\eta_{|\nu}
	-\frac{5\kappa^2}{12}h_{\mu\nu}\eta^{|\alpha}\eta_{|\alpha}
	-\Lambda(\eta)h_{\mu\nu}
	+\frac{\kappa^2}{\ell}T_{\mu\nu}
	-\overset{(1)}{E}_{\mu\nu} \ , 
	\label{MW1}\\
\Box\eta&=&V'+
	\frac{\kappa^2}{3}\sigma_0\sigma'
	-\frac{\kappa^2}{12}\sigma'T
	-\overset{(1)}{\Phi}{}_2 \ ,
	\label{MW2}
\end{eqnarray}
where $\Lambda(\eta)=\kappa^2 V/2 + \kappa^4 \sigma_0\tilde\sigma/6$.
As we have solved the bulk equations, we can
write $E_{\mu\nu}$ or $\Phi_2$ explicitly in terms of
4-dimensional quantities. At first order, we obtain
\begin{eqnarray}
\overset{(1)}{E}{}^\mu{}_\nu = {2\over\ell}\chi^\mu{}_\nu 
	-{\kappa^2\over 3}\left(\eta^{|\mu} \eta_{|\nu}
	-{1\over 4}\delta^\mu_\nu \eta^{|\alpha}\eta_{|\alpha}
  	\right) 
  	\label{1:weyl}\ .
\end{eqnarray}
In cosmology, this Weyl tensor leads to  
$\frac{2}{\ell}\chi^{0}{}_{0}+\frac{\kappa^2}{4}\dot{\eta}^2$ which
coincides with the one used by Langlois and Sasaki~\cite{langlois}. 
Substituting Eq.~(\ref{1:weyl}) into Eq.~(\ref{MW1}), 
we see the geometrical approach agrees with our result (\ref{1:einstein}).
Remarkably, the energy-momentum tensor for the field $\eta$
 is transformed to the standard form thanks to  $E_{\mu\nu}$. 
We also obtain
\begin{eqnarray}
\overset{(1)}{\Phi}_2= \Box \eta +{4\over \ell} C \ .
\label{1:phi2}
\end{eqnarray}
Substituting Eq.~(\ref{1:phi2}) into Eq.~(\ref{MW2}), the geometrical 
approach agrees 
with  Eq.~(\ref{1:KG}). 
 It explains the factor $1/2$ in the effective potential (\ref{1:effpt}).

\subsection{Two-brane Model}

Now we can apply our results in the previous section to 
two-brane system 
 and write down effective equations on each brane.
 
 Our system is described by the action 
\begin{eqnarray}
S&=&{1\over 2 \kappa^2}\int d^5 x \sqrt{-g}\left({\cal R}
+{12\over \ell^2}\right)-\sum_{i=\oplus,\ominus}\overset{i}{\sigma}(\varphi) 
\int d^4 x \sqrt{-g^{i\mathrm{\hbox{-}brane}}}
\nonumber\\
&&+\sum_{i=\oplus,\ominus} \int d^4 x \sqrt{-g^{i\mathrm{\hbox{-}brane}}}
\,{\cal L}_{\rm matter}^i \ ,
	\label{5D:action}
\end{eqnarray}
where  $g^{i\mathrm{\hbox{-}brane}}_{\mu\nu}$ 
and $\overset{i}{\sigma}$ are  the induced metric 
 and the brane tension on the $i$-brane, respectively. 
 We consider an $S_1/Z_2$ orbifold spacetime with the two branes 
as the fixed points. In the first Randall-Sundrum (RS1) model, 
the two flat 3-branes are embedded in AdS$_5$ 
 and the brane tensions given by
$\overset{\oplus}{\sigma}=6/(\kappa^2\ell)$ and 
$\overset{\ominus}{\sigma}=-6/(\kappa^2\ell)$. 
 
The point is the fact that the equations of motion on each brane
take the same form if we use the induced metric on each brane~\cite{kanno2}
 (see also \cite{wiseman} for other approaches). 
 The effective Einstein equations on each positive ($\oplus$)
and negative ($\ominus$) tension brane at low-energies yield
\begin{eqnarray}
G^\mu{}_\nu (h) &=& \kappa^2 \left( 
	\eta^{|\mu}\eta_{|\nu}
        -{1\over 2}\delta^\mu_\nu \eta^{|\alpha} \eta_{|\alpha}
        -{1\over 2}\delta^\mu_\nu V\right) 
        -{2\over \ell} \chi^\mu{}_\nu
        +\frac{\kappa^2}{\ell}\left(
        \overset{\oplus}{T}{}^{\mu}{}_{\nu}
        -\delta^{\mu}_{\nu}\overset{\oplus}{\tilde\sigma}\right)
        \label{p:einstein}    \ , \\
G^\mu{}_\nu (f) &=& \kappa^2 \left( 
        \eta^{;\mu} \eta_{;\nu}
        -{1\over 2}\delta^\mu_\nu \eta^{;\alpha} \eta_{;\alpha}
        -{1\over 2}\delta^\mu_\nu V\right)
        -{2\over \ell} {\chi^\mu{}_\nu \over \Omega^4}
        -\frac{\kappa^2}{\ell}\left( 
        \overset{\ominus}{T}{}^{\mu}{}_{\nu}
        -\delta^{\mu}_{\nu}\overset{\ominus}{\tilde\sigma}\right)
        \label{n:einstein-1}  \ .
\end{eqnarray}
where $f_{\mu\nu}$ is the induced metric on the negative tension brane 
and $;$ denotes the covariant derivative with respect to $f_{\mu\nu}$. 
 When we set the position of the positive tension brane at $y=0$, 
  that of the negative tension brane $\bar{y}$ in general depends on
 $x^\mu$, i.e. $\bar{y} = \bar{y} (x^\mu)$. Hence,  
 the warp factor at the negative tension brane 
 $\Omega (x^\mu)\equiv b(\bar{y}(x)) $ also depends on $x^\mu$.
  Because the metric always comes into equations
 with  derivatives, the zeroth order relation is enough in this 
 first order discussion. Hence, the metric on the
 positive tension brane is related to the metric on the 
 negative tension brane as $f_{\mu\nu} = \Omega^2 h_{\mu\nu}$.
   
Although Eqs.~(\ref{p:einstein}) and (\ref{n:einstein-1}) are non-local 
individually, with undetermined $\chi^{\mu}{}_{\nu}$, one can combine both
equations to reduce them to local equations for each brane. 
 Eq. (\ref{n:einstein-1}) can be rewritten 
using the induced metric on the positive tension brane as
\begin{eqnarray}
&&{1\over \Omega^2}\left[G^{\mu}{}_{\nu} (h) 
	-2\left(\log \Omega \right)^{|\mu}{}_{|\nu}
	+2\delta^\mu_\nu \left(\log \Omega \right)^{|\alpha}{}_{|\alpha}
 	+2\left(\log \Omega \right)^{|\mu} 
 	\left(\log \Omega \right)_{|\nu}
 	+\delta^\mu_\nu \left(\log \Omega \right)^{|\alpha} 
 	\left(\log \Omega \right)_{|\alpha}  \right] \nonumber \\
&& \qquad 
	= {\kappa^2 \over \Omega^2} \left( 
        \eta^{|\mu} \eta_{|\nu}
        -{1\over 2}\delta^\mu_\nu \eta^\alpha \eta_\alpha \right)
        -{\kappa^2 \over 2} V \delta^\mu_\nu 
        -{2\over \ell} {\chi^\mu{}_\nu \over \Omega^4}
        -\frac{\kappa^2}{\ell}\left(
        \overset{\ominus}{T}{}^{\mu}{}_{\nu}
        -\delta^{\mu}_{\nu}\tilde\sigma\right) \ .
        \label{n:einstein-2}
\end{eqnarray}
We can therefore easily eliminate $\chi^{\mu}{}_{\nu}$ from 
Eqs.~(\ref{p:einstein}) and (\ref{n:einstein-2}), since 
$\chi^{\mu}{}_{\nu}$ appears only algebraically.
Eliminating $\chi^\mu{}_\nu$ from both Eqs.~(\ref{p:einstein}) and 
(\ref{n:einstein-2}), we obtain 
\begin{eqnarray}
G^\mu{}_\nu &=&
	\frac{\kappa^2}{\ell\Psi}
	\overset{\oplus}{T}{}^{\mu}{}_{\nu}
	+\frac{\kappa^2(1-\Psi)^2}{\ell\Psi}
	\overset{\ominus}{T}{}^{\mu}{}_{\nu}
        \nonumber\\
&&
	+{1\over \Psi}\left[ \Psi^{|\mu}{}_{|\nu} 
     	- \delta^\mu_\nu \Psi^{|\alpha}{}_{|\alpha} 
     	+{3\over 2}{1 \over 1-\Psi} \left(
     	\Psi^{|\mu} \Psi_{|\nu} -{1\over 2}\delta^\mu_\nu
     	\Psi^{|\alpha} \Psi_{|\alpha} \right) \right] \nonumber\\
&& 
	+\kappa^2\left( 
        \eta^{|\mu} \eta_{|\nu}
        -\frac{1}{2}\delta^\mu_\nu\eta^{|\alpha}\eta_{|\alpha}
	-\delta^\mu_\nu V_{\rm eff}
        \right)
        \label{QST1}
         \ .
\end{eqnarray}
where we defined a new field $\Psi=1-\Omega^2$ 
which we refer to by the name ``radion",
 and the effective potential takes the form
\begin{eqnarray}
V_{\rm eff}=\frac{1}{\ell}\left[~
	\frac{1}{\Psi}\overset{\oplus}{\tilde\sigma'}
	+\frac{(1-\Psi)^2}{\Psi}\overset{\ominus}{\tilde\sigma'}~\right]
	+\frac{2-\Psi}{2}V \ .
\end{eqnarray}
 The bulk scalar field induces the energy-momentum
 tensor of the conventional 4-dimensional scalar field with 
 the effective potential which depends on the radion. 

We can also determine the dark radiation $\chi^{\mu}{}_{\nu}$ by eliminating 
$G^{\mu}{}_{\nu}(h)$ from Eqs.~(\ref{p:einstein}) and
(\ref{n:einstein-2}),
\begin{eqnarray}
{2\over \ell}\chi^\mu{}_{\nu}
	&=&-{1\over \Psi}\left[ \Psi^{|\mu}{}_{|\nu} 
     	- \delta^\mu_\nu \Psi^{|\alpha}{}_{|\alpha} 
     	+{3\over 2}{1 \over 1-\Psi} \left(
     	\Psi^{|\mu} \Psi_{|\nu} -{1\over 2}\delta^\mu_\nu
     	\Psi^{|\alpha} \Psi_{|\alpha} \right) \right] 
     	\nonumber\\
&&
     	+{\kappa^2 \over 2}(1-\Psi)\delta^\mu_\nu V
     	-\frac{\kappa^2}{\ell}\frac{1-\Psi}{\Psi}
     	\left[\overset{\oplus}{T}{}^{\mu}{}_{\nu}
     	-\delta^{\mu}_{\nu}\overset{\oplus}{\tilde\sigma}
     	+\left(1-\Psi\right)
     	\left(\overset{\ominus}{T}{}^{\mu}{}_{\nu}
     	-\delta^{\mu}_{\nu}\overset{\ominus}{\tilde\sigma}\right)\right] \ .
     	\label{chi2}
\end{eqnarray}
Due to the property $\chi^\mu{}_\mu =0$, we have
\begin{eqnarray}
\Box\Psi&=&
	\frac{\kappa^2}{3\ell}(1-\Psi)\left[
  	\overset{\oplus}{T}
  	+(1-\Psi)\overset{\ominus}{T}\right]
	-{1\over 2(1-\Psi)} \Psi^{|\alpha} \Psi_{|\alpha} 
	\nonumber\\
&&\qquad
  	-{2\kappa^2\over 3}\Psi(1-\Psi)V
  	-\frac{4\kappa^2}{3\ell}(1-\Psi)
  	\left[\overset{\oplus}{\tilde\sigma}
  	+(1-\Psi)\overset{\ominus}{\tilde\sigma}
  	\right] 
  	\label{QST2}\ .
\end{eqnarray}
Note that Eqs.~(\ref{QST1}) and (\ref{QST2}) are derived from 
a scalar-tensor type theory coupled to the  additional scalar field.

Similarly, the equations for the scalar field on branes become
\begin{eqnarray}
&&\Box_h\eta-{V^\prime\over 2}+{2\over\ell}C
	=\frac{1}{\ell}\overset{\oplus}{\tilde\sigma'} \ ,
	\label{p:KG}\\
&&\Box_f\eta-{V^\prime\over 2}+{2\over\ell}{C\over\Omega^4}
	=-\frac{1}{\ell}\overset{\ominus}{\tilde\sigma'} \ ,
	\label{n:KG}
\end{eqnarray}
where the subscripts refer to the induced metric on each brane. 
Notice that the scalar field takes the same value for both branes 
at this order.
Eq.(\ref{n:KG}) can be written using the induced metric on the 
positive tension brane as
\begin{eqnarray}
{1\over \Omega^2} \Box_h \eta 
	+{1\over \Omega^2}( \log \Omega^2 )^{|\mu} \eta_{|\mu}
	-{V^\prime \over 2} + {2\over \ell} {C\over \Omega^4} 
	=-\frac{1}{\ell}\overset{\ominus}{\tilde\sigma'} \ .
	\label{n:KG-2}
\end{eqnarray}
Eliminating the dark source $C$ from these Eqs.~(\ref{p:KG}) 
and (\ref{n:KG-2}), we find the equation for the scalar field 
takes the form
\begin{eqnarray}
\Box_h\eta-V'_{\rm eff}
	=-\frac{\Psi^{|\mu}}{\Psi}\eta_{|\mu} \ .
	\label{p:KG-2}
\end{eqnarray}
Notice that the radion acts as a source for $\eta$. And we can also 
get the dark source as
\begin{eqnarray}
{2\over \ell}C=-{V^\prime\over 2}(1-\Psi)
	+{\Psi^{|\mu}\over\Psi}\eta_{|\mu}
	-\frac{1-\Psi}{\ell\Psi}\left[
	\overset{\oplus}{\tilde\sigma'}
	+(1-\Psi)\overset{\ominus}{\tilde\sigma'}\right] \ .
	\label{C2}
\end{eqnarray}
Interestingly, $\chi^{\mu}_{\nu}$ and $C$ vanishes in the single brane 
limit, $\Psi \rightarrow 1$ as can be seen from (\ref{chi2}) and (\ref{C2}).

Now the effective action for the positive tension brane which
gives Eqs.~(\ref{QST1}), (\ref{QST2}) and (\ref{p:KG-2}) can be read off as
\begin{eqnarray}
S&=&{\ell\over 2\kappa^2}\int d^4x\sqrt{-h}\left[
	\Psi R
	-\frac{3}{2(1-\Psi)}\Psi^{|\alpha} \Psi_{|\alpha} 
	-\kappa^2\Psi\left(\eta^{|\alpha}\eta_{|\alpha}
	+2V_{\rm eff}\right)
	\right]  
	\nonumber\\
&&
	+\int d^4x\sqrt{-h}~
	\overset{\oplus}{\cal L}
	+\int d^4x\sqrt{-h}~(1-\Psi)^2
	\overset{\ominus}{\cal L}  \ ,
\end{eqnarray}
where the last two terms represent actions for the matter on each brane.  
Thus, we found the radion field couples with the induced metric and 
the induced scalar field on the brane non-trivially. 
 Surprisingly, at this order, the nonlocality of 
 $\chi_{\mu\nu}$ and $C$
 are eliminated by the radion. 
 As this is a close system, we can analyze a primordial spectrum
 to predict the  cosmic background fluctuation spectrum~\cite{soda2}.

\section{Kaluza-Klein Corrections }

In the previous section, we showed that the leading order results
in our iteration 
scheme yield the usual Einstein-scalar gravity in the case of the 
single-brane system.  
In the non-dilatonic braneworld, it is now well known that Kaluza-Klein (KK) 
modes generate a correction to the Newtonian force. 
Therefore, we  expect that the next order calculation in our iteration scheme 
 gives a similar correction to the usual Einstein-scalar gravity.
In this section, therefore, we consider  corrections coming from KK modes.
We derive solutions at this order in Appendix B.  

Substituting solutions (\ref{2:trace}) and (\ref{2:traceless}) 
into the junction condition, 
 we obtain the effective Einstein equation
\begin{eqnarray}
G^{\mu}{}_{\nu}&=&
	\frac{\kappa^2}{\ell}T^{\mu}{}_{\nu}
	+\kappa^2\left[
	\eta^{|\mu}\eta_{|\nu}
	-\frac{1}{2}\delta^{\mu}_{\nu}
	\eta^{|\alpha}\eta_{|\alpha}
	-\delta^{\mu}_{\nu}V_{\rm eff}\right]
	\nonumber\\
&&	+\frac{\ell^2}{2}
	{\cal S}^{\mu}{}_{\nu}
	-\frac{\ell^2}{4}\kappa^2\left(V''
	-\frac{\kappa^2}{3}V\right){\cal U}^{\mu}{}_{\nu}
	-\frac{\ell^2}{12} \kappa^2 {\cal Q}^{\mu}{}_{\nu}
	+{\kappa^2 \over \ell} \tau^{\mu}{}_{\nu}  \ ,
	\label{2:einstein}
\end{eqnarray}
where the effective potential at this order is defined by 
\begin{eqnarray}
V_{\rm eff}=\frac{1}{\ell}\tilde\sigma
	+\frac{1}{2}V
	+\frac{\ell^2\kappa^2}{48}V^2
	-\frac{\ell^2}{64}V^{\prime 2} 
	\label{2:effpt}\ .
\end{eqnarray}
Here ${\cal S}^{\mu}{}_{\nu},\ {\cal U}^{\mu}{}_{\nu}$ and 
${\cal Q}^{\mu}{}_{\nu}$ are the quantities for which one can write 
the Lagrangian explicitly in terms of $R^{\mu}{}_{\nu}$ and $\eta$. 
We present the explicit expressions for these quantities in Appendix A. 
On the other hand, we cannot write the local Lagrangian for
 $\tau^{\mu}{}_{\nu}$  given by
\begin{eqnarray}	
\tau^{\mu}{}_{\nu}&=&
	-\frac{\ell^3}{12\kappa^2 }\left[
	-RR^{\mu}{}_{\nu}
	+\kappa^2R^{\mu}{}_{\nu}\eta^{|\alpha}\eta_{|\alpha}
	+\kappa^2R~\eta^{|\mu}\eta_{|\nu}
	\right.
	\nonumber\\
&&	\left.
	+\frac{1}{2}\delta^{\mu}_{\nu}\left(
	R^2-\frac{3}{2}R^{\alpha}{}_{\beta}R^{\beta}{}_{\alpha}
	-2\kappa^2R\eta^{|\alpha}\eta_{|\alpha}
	+3\kappa^2R^{\alpha\beta}\eta_{|\alpha}\eta_{|\beta}
	-\frac{3}{2}\kappa^2(\Box\eta)^2
	-\kappa^4(\eta^{|\alpha}\eta_{|\alpha})^2
	\right)\right] \nonumber\\
&&	-\frac{\ell^3}{4\kappa^2 }\left[
	2\kappa^2{\cal J}^{\mu}{}_{\nu}
	-\frac{2}{3}\kappa^2{\cal L}^{\mu}{}_{\nu}
	-\kappa^2{\cal P}^{\mu}{}_{\nu}
	-\frac{5}{6}\kappa^4{\cal M}^{\mu}{}_{\nu}
	\right]
	-\frac{2}{\kappa^2}\overset{(2)}{\chi}{}^{\mu}{}_{\nu} \ .
\end{eqnarray}
Here ${\cal J}^{\mu}{}_{\nu}, {\cal L}^{\mu}{}_{\nu}, {\cal P}^{\mu}{}_{\nu}$
and ${\cal M}^{\mu}{}_{\nu}$ are also defined in Appendix A.
Interestingly,  the trace of $\tau^{\mu}{}_{\nu}$ becomes
the following form 
\begin{eqnarray}
\tau = \frac{\ell^3}{4\kappa^2}\left[
	R^{\alpha\beta}R_{\alpha\beta}-\frac{1}{3}R^2
	-2\kappa^2R^{\alpha\beta}\eta_{|\alpha}\eta_{|\beta}
	+\frac{2}{3}\kappa^2R~\eta^{|\alpha}\eta_{|\alpha}
	+\kappa^2(\Box\eta)^2
	+\frac{2}{3}\kappa^4(\eta^{|\alpha}\eta_{|\alpha})^2\right] \ .
\end{eqnarray}
This is the more familiar form of trace-anomaly~\cite{Nojiri:2000ja}.
 It is here AdS/CFT correspondence comes into the braneworld.
 Namely, $\tau_{\mu\nu}$ can be interpreted as the energy-momentum
 tensor of the CFT matter. Note that there is an ambuiguity 
 to separate the action into CFT part and other part. However, the 
 same problem exists even  in the non-dilatonic case~\cite{kanno1}.
 Despite of this defect, it is generally believed that AdS/CFT interpretation 
 is useful at least qualitatively. Our finding must be also beneficial 
 for understanding of the non-linear physics in the braneworld.   
  
Using the Eq.~(\ref{2:scalar}) in Appendix B, the junction condition 
for the scalar field leads to 
\begin{eqnarray}
&&\Box\eta-V'_{\rm eff}
	+\frac{\ell^2}{24}V'\left(R+\kappa^2\eta^{|\alpha}\eta_{|\alpha}\right)
	-\frac{\ell^2}{4}\left(V''-\frac{\kappa^2}{3}V\right)\Box\eta
	-\frac{\ell^2}{8}V'''\eta^{|\alpha}\eta_{|\alpha}
	=J   \ .
	\label{2:KG}
\end{eqnarray}
Here we have defined the  source $J$ with KK corrections as
\begin{eqnarray}
J&=&\frac{\ell^2}{4}\left(
	\Box^2\eta
	+2R^{|\alpha\beta}\eta_{|\alpha\beta}
	-\frac{1}{3}R\Box\eta
	+\frac{1}{3}R^{|\alpha}\eta_{|\alpha}
	-\frac{8}{3}\kappa^2\eta^{|\alpha}\eta^{|\beta}\eta_{|\alpha\beta}
	-\frac{5}{3}\kappa^2\eta^{|\alpha}\eta_{|\alpha}\Box\eta\right)
	-\frac{2}{\ell}\overset{(2)}{C} \ . 
\end{eqnarray}
Using the Bianchi identities (or momentum constraint),
we obtain the constraint
on $\tau_{\mu\nu}$ and the source $J$ 
\begin{eqnarray}	
\tau_{\mu\nu}{}^{|\nu} =-\ell J~\eta_{|\mu}  \ .
\label{CFT:conservation} 
\end{eqnarray}
This tells us that CFT matter cannot be confined to the brane in contrast to
the non-dilatonic braneworld. 

Since the Lagrangian for each parts of the effective Einstein equations is 
 known (see Appendix A), we can now deduce the effective action   
\begin{eqnarray}
S&=&\frac{\ell}{2\kappa^2}\int d^4x\sqrt{-h}\left[
	\left(1+\frac{\ell^2}{12}\kappa^2V\right)R
	-\kappa^2\left(
	1+\frac{\ell^2}{12}\kappa^2V-\frac{\ell^2}{4}V''\right)
	\eta^{|\alpha}\eta_{|\alpha}
	-2\kappa^2V_{\rm eff}
	\right.\nonumber\\
&&\left.\hspace{3cm}
	-\frac{\ell^2}{4}\left(
	R^{\alpha\beta}R_{\alpha\beta}
	-\frac{1}{3}R^2
	\right)\right]
	+\int d^4 x \sqrt{-h}~ 
	{\cal L}_{\rm matter}
	+S_{\rm CFT} \ ,
\end{eqnarray}
where the last term comes from 
 the energy-momentum tensor of CFT matter $\tau_{\mu\nu}$. 
Although it is difficult to obtain the explicit
form  of the non-local Lagrangian for $\tau_{\mu\nu}$, we know variations 
 with respect to $h_{\mu\nu}$ and $\eta$. By definition, we have 
\begin{eqnarray}
\frac{-2}{\sqrt{-h}}\frac{\delta S_{\rm CFT}}{\delta h^{\mu\nu}}&=& 
	\tau_{\mu\nu} \ .
\end{eqnarray}
The variation with respect to $\eta$ yields the equation of motion for the 
scalar field. Comparing with the result from junction condition
for the scalar field, we get the following relation
\begin{eqnarray}
\frac{1}{\sqrt{-h}}\frac{\delta S_{\rm CFT}}{\delta\eta}&=&
	-\ell~J \ .
\end{eqnarray}
This means the CFT matter couples with the scalar field. Note that 
Eq.(\ref{CFT:conservation}) guarantees the
diffeomorphism invariance of $S_{CFT}$. 
 Here, we do not intend to insist that we have determined the constant
 of integration which is nonlocal in general, rather we have transformed
 this unknown quantity to some known nonlocal CFT matter. Of course,
 which kind of CFT matter should be chosen is remained to be solved. 
 Admittedly, this is the defect of our approach. We just hope the 
 proper consideration of the boundary condition or more fundamental
 development of the superstring theory change the situation.

Let us see the relation of our results to the geometric approach
at this order as well.
If we rewrite Eqs.~(\ref{2:einstein}) and 
(\ref{2:KG}) using the Weyl tensor $E_{\mu\nu}$ and $\Phi_2$ 
 whose explicit forms at second order can be found in Eqs.~(\ref{2:weyl}) 
 and 
 (\ref{2:phi2})  in Appendix B  
and reduce them using Eqs.~(\ref{1:einstein}) and 
(\ref{1:KG}) at first order, finally we obtain
\begin{eqnarray}
&&G^{\mu}{}_{\nu}=
	\frac{2\kappa^2}{3}\eta^{|\mu}\eta_{|\nu}
	-\frac{5\kappa^2}{12}\delta^{\mu}_{\nu}\eta^{|\alpha}\eta_{|\alpha}
	-\delta^{\mu}_{\nu}\Lambda(\eta)
	+\frac{\kappa^2}{\ell}T^{\mu}{}_{\nu}
	+\frac{\kappa^4}{6}\tilde\sigma T^{\mu}{}_{\nu}
	+\kappa^4\pi^{\mu}{}_{\nu}
	-\overset{(1)}{E}{}^{\mu}{}_{\nu}
	-\overset{(2)}{E}{}^{\mu}{}_{\nu} \ ,
	\label{2:MW1}\ \\
&&\Box\eta=V'+\frac{\kappa^2}{3}\sigma_0\tilde\sigma'
	+\frac{\kappa^2}{3}\tilde\sigma\tilde\sigma'
	-\frac{\kappa^2}{12}\tilde\sigma'T
	-\overset{(1)}{\Phi}{}_2
	-\overset{(2)}{\Phi}{}_2 
	\label{2:MW2}\ ,
\end{eqnarray}
where
\begin{eqnarray}
&&\Lambda(\eta)=\frac{\kappa^2}{\ell}\tilde\sigma
	+\frac{\kappa^2}{2}V
	+\frac{\kappa^4}{12}\tilde\sigma^2
	-\frac{\kappa^2}{16}\tilde\sigma^{\prime 2} \ ,
	\label{cosmconst}\\
&&\pi_{\mu\nu}=-{1\over 4}T_\mu^\lambda T_{\lambda\nu} 
  +{1\over 12}TT_{\mu\nu} + {1\over 8} g_{\mu\nu} \left( 
   T^{\alpha\beta} T_{\alpha\beta} -{1\over 3} T^2 \right) \ .
\end{eqnarray}
Here $\tau_{\mu\nu}$ is absorbed in Weyl tensor.
These are nothing but equations obtained by Maeda and Wands~\cite{maeda}.

At this point, we give a rather general result as a byproduct of our analysis. 
Indeed, our result suggests the following generalization of arguments 
advocated by Langlois and Sasaki~\cite{langlois}. 
Let us rewrite equations obtained by 
 Maeda and Wands as 
\begin{eqnarray}
           G_{\mu\nu}&=&
	\kappa^2 \partial_\mu\varphi\partial_\nu\varphi
	-{\kappa^2\over 2}g_{\mu\nu} 
	\partial^\alpha\varphi\partial_\alpha\varphi
	-\Lambda(\varphi)g_{\mu\nu} 
	+ X_{\mu\nu} 
  	   \ , \\
\Box\varphi &=& {1\over \kappa^2} \Lambda' (\varphi)+\tilde{J}  \ ,  	
\end{eqnarray}
where , following Langlois and Sasaki, we have defined
\begin{eqnarray}
X_{\mu\nu}  &=& {\kappa^4\over 6}\sigma(\varphi)T_{\mu\nu}
	+\kappa^4\pi_{\mu\nu}-E_{\mu\nu}
	-{\kappa^2\over 3}
	\left(\partial_\mu \varphi \partial_\nu \varphi
	-{1\over 4} g_{\mu\nu} 
	\partial^\alpha \varphi \partial_\alpha \varphi \right)
	             \ , \\
 \tilde{J} &=&  \frac{1}{8} \sigma^\prime \sigma^{\prime\prime}
	-\frac{\kappa^2}{24} \sigma' T
	+{1\over 2} \Box \varphi
	- {\Phi_2 \over 2}  \ .
\end{eqnarray}	
Here, we have transformed the non-conventional kinetic term into
 the standard form. 
Thus, $\Lambda$ can be interpreted as the potential functional
 for the scalar field $\varphi$. 
Bianchi identity implies
\begin{eqnarray}
  \nabla_\nu X^\nu{}_{\mu } = - \kappa^2 \tilde{J} \nabla_\mu \varphi \ .
  \label{DR:conservation}
\end{eqnarray}
In the absence of the matter $T_{\mu\nu}$ on the brane, we have
 $X^\mu{}_\mu =0$, namely, we can interpret $X^\mu{}_\nu$ as the
 energy momentum tensor for the dark radiation. In this case,
 Eq.~(\ref{DR:conservation}) describes the leak of the dark radiation 
 due to the bulk scalar field. 
 It should be stressed that no approximation is employed 
 in the above argument.

%
\section{Conclusion}

 We derived the non-linear low energy effective action for the 
 dilatonic braneworld.
 We considered the bulk scalar field  with a nontrivial potential and
  the brane tension coupled to the bulk scalar field.
 Provided that the energy density of the matter on the brane is less than 
  the brane tension and the potential energy in the bulk is 
 less than the bulk vacuum energy, we formulated the systematic low energy 
 expansion scheme to solve the bulk equations 
 and derived the effective 4-dimensional action for both the single-brane 
 model and  the two-brane model through the junction conditions. 
 
  In the case of the single-brane model, the effective theory is described 
 by the  Einstein-scalar theory  with the dark radiation. 
 Remarkably, the dark radiation is not conserved 
 in general due to the coupling to the bulk scalar. We also clarified the
 role of $E_{\mu\nu}$ in the dilatonic braneworld, namely, enigmatic
  kinetic terms in the formula of the geometric equations reduce
 to that of the conventional scalar field thanks to $E_{\mu\nu}$. 
 A  factor $1/2$ in the effective potential (\ref{2:effpt})
   is also naturally explained by $\Phi_2$.  
 Admittedly, the dark radiation and the dark source can not be determined
 without imposing the boundary conditions at AdS horizon. 
 If we choose the regularity  condition at the horizon, these
 terms must vanish. This is the reason why the conventional Einstein
 theory can be recovered at the low energy~\cite{others1}. 
 
 In the case of the two-brane model, junction conditions at both branes
 give the boundary conditions.
 Then, the constant of integration is determined completely.
 As a result, the effective theory reduces to the scalar-tensor
 theory with the non-trivial coupling between the radion and the bulk scalar
 field.  It turns out that $\chi_{\mu\nu}$ and $C$ becomes zero 
 when two branes get separated  infinitely. 
 However, we should be careful to treat this limit, 
 because  next order corrections  might diverge in this limit
 as is suggested by analysis of the linearized  gravity~\cite{kanno2}. 
 
 As for the single-brane model, we also constructed the effective action
 with  Kaluza-Klein corrections and revealed the role of the AdS/CFT
  correspondence.  In particular, it was shown that CFT matter would not be 
  confined to the braneworld in the presence of the bulk scalar field. 
 The relation between our analysis and the geometrical projection method was
 also clarified at this order. In particular, we have shown that the quadratic
 part of the energy-momentum tensor $\pi_{\mu\nu}$ can be reproduced
 from our effective action. Hence, by analyzing our effective action,
 we can investigate both  high-energy effects and  effects of 
 $E_{\mu\nu}$. It should be noted that we did not determine the constant
 of integration which is nonlocal in general. Instead we transformed
 the issue to another form, namely, choosing the approproate CFT
 matter suitable for the physical situation. 
 
 Moreover, we generalized the conjecture of Langlois and Sasaki
 which is explicitly advocated only in the homogeneous model. 
 The key observation is that the enigmatic kinetic part of
 the scalar field can be transformed to the standard form by subtracting
 the traceless combination of $\partial_\mu \varphi \partial_\nu \varphi$ 
 which is absorbed into the energy-momentum tensor of dark radiation.

\begin{acknowledgments}
We would like to thank Misao Sasaki for useful discussions. 
This work was supported in part by  Monbukagakusho Grant-in-Aid No.14540258. 
\end{acknowledgments}


\appendix

\section{Useful Formula}
In this Appendix, we give  possible local functionals
in terms of $R_{\mu\nu}$ and $\eta$  up to
  second order in our low-energy expansion.
   
First of all, we consider the following variation
\begin{eqnarray}
\delta\int d^4x\sqrt{-h}~L(h_{\mu\nu}, \eta)
	=\int d^4x\sqrt{-h}\left[
	\delta h^{\mu\nu}X_{\mu\nu}
	+\delta\eta\frac{\delta L}{\delta\eta}
	\right] \ .
\end{eqnarray}
Provided the invariance under the diffeomorphism  
$\delta h_{\mu\nu} = - \xi_{\mu |\nu}-\xi_{\nu |\mu}$ and
$\delta \eta = -\xi^\mu \eta_{|\mu}$, we get the identity
\begin{eqnarray}
X_{\mu\nu}{}^{|\nu}
	+\frac{1}{2}\frac{\delta L}{\delta\eta}\eta_{|\mu}
	=0 \ .
	\label{conservation}
\end{eqnarray}
Conversely, this identity guarantees the diffeomorphism invariance of the system.
This fact play an important role in the discussion related to the AdS/CFT
 correspondence. 

Besides the above theoretical importance, the above identity reduces the
necessary calculations considerably.  
Hence, we apply the above identity to the following tensors.

The relevant local functional  with fourth order derivatives 
which consists of $R_{\mu\nu}$ 
is given by 
\begin{eqnarray}
&&\delta\int d^4 x ~\sqrt{-h}~{1\over 2}\left[ 
	R^{\alpha\beta}R_{\alpha\beta}
	-{1\over 3}R^2 \right] 
	=\int d^4 x \sqrt{-h}~{\cal S}_{\mu\nu}~
        \delta h^{\mu\nu} \ ,
\end{eqnarray}
where
\begin{eqnarray}
&&{\cal S}_{\mu\nu}=R_{\mu\alpha}R^{\alpha}{}_{\nu}
	-{1\over 3}RR_{\mu\nu} 
	-{1\over 4}h_{\mu\nu}\left(R^{\alpha\beta}R_{\alpha\beta}
	-{1\over 3}R^2\right)  \nonumber \\ 
&&\qquad\qquad\qquad
	-{1\over 2}\left( R_{\mu}{}^{\alpha}{}_{|\nu\alpha}
	+R_{\nu}{}^{\alpha}{}_{|\mu\alpha}  
	-{2\over 3}R_{|\mu\nu}
	-\Box R_{\mu\nu} 
	+{1\over 6}h_{\mu\nu}\Box R \right)\ .
\end{eqnarray}
The local functionals with fourth order derivatives 
which consist of $R_{\mu\nu}$ and $\eta$ are 
\begin{eqnarray}
&&\delta\int d^4 x \sqrt{-h}\left(R~\eta^{|\alpha}\eta_{|\alpha}\right)
	=\int d^4 x \sqrt{-h}~{\cal L}_{\mu\nu}~
                         \delta h^{\mu\nu} \ , \\
&&\delta\int d^4 x \sqrt{-h}
	\left(R^{\alpha\beta}\eta_{|\alpha}\eta_{|\beta}\right)
	=\int d^4 x \sqrt{-h}~{\cal J}_{\mu\nu}~
        \delta h^{\mu\nu} \ ,
\end{eqnarray}
where
\begin{eqnarray}
&&
{\cal L}_{\mu\nu}=
	-\frac{1}{2}h_{\mu\nu}R~\eta^{|\alpha}\eta_{|\alpha}
	+R_{\mu\nu}\eta^{|\alpha}\eta_{|\alpha}
	-(\eta^{|\alpha}\eta_{|\alpha})_{|\mu\nu}
	+h_{\mu\nu}\Box(\eta^{|\alpha}\eta_{|\alpha})
	+R~\eta_{|\mu}\eta_{|\nu} \ ,\\
&&
{\cal J}_{\mu\nu}=
	-\frac{1}{2}h_{\mu\nu}R^{\alpha\beta}
	\eta_{|\alpha}\eta_{|\beta}
	+R_{\mu}{}^{\alpha}\eta_{|\nu}\eta_{|\alpha}
	+R_{\nu}{}^{\alpha}\eta_{|\alpha}\eta_{|\mu}
	\nonumber\\
&&\qquad\qquad\qquad
	-\frac{1}{2}(\eta_{|\mu}\eta^{|\alpha})_{|\nu\alpha}
	-\frac{1}{2}(\eta_{|\nu}\eta^{|\alpha})_{|\mu\alpha}
	+\frac{1}{2}\Box(\eta_{|\mu}\eta_{|\nu})
	+\frac{1}{2}h_{\mu\nu}
	(\eta^{|\alpha}\eta^{|\beta})_{|\alpha\beta} \ .
\end{eqnarray}
The local functionals with fourth order derivatives 
which consist of $\eta$  are
\begin{eqnarray}
&&\delta\int d^4 x \sqrt{-h}~(\Box\eta)^2
	=\int d^4 x \sqrt{-h}~{\cal P}_{\mu\nu}~
                         \delta h^{\mu\nu}     \ ,\\
&&\delta\int d^4 x \sqrt{-h}~(\eta^{|\alpha}\eta_{|\alpha})^2
	=\int d^4 x \sqrt{-h}~{\cal M}_{\mu\nu}~
                         \delta h^{\mu\nu}     \ ,
\end{eqnarray}
where
\begin{eqnarray}
&&
{\cal M}_{\mu\nu}=
	-\frac{1}{2}h_{\mu\nu}(\eta^{|\alpha}\eta_{|\alpha})^2
	+2\eta_{|\mu}\eta_{|\nu}\eta^{|\alpha}\eta_{|\alpha} \ ,\\
&&
{\cal P}_{\mu\nu}=
	\frac{1}{2}h_{\mu\nu}(\Box\eta)^2
	-(\Box\eta)_{|\mu}\eta_{\nu}
	-(\Box\eta)_{|\nu}\eta_{\mu}
	+h_{\mu\nu}(\Box\eta)^{|\alpha}\eta_{\alpha} \ .
\end{eqnarray}
The local functionals with second order derivatives are
\begin{eqnarray}
&&\delta\int d^4 x \sqrt{-h}~\eta^{|\alpha}\eta_{|\alpha}
	=\int d^4 x \sqrt{-h}~{\cal U}_{\mu\nu}~
        \delta h^{\mu\nu}     \ ,\\
&&\delta\int d^4 x \sqrt{-h}~R~V(\eta)
	=\int d^4 x \sqrt{-h}~{\cal Q}_{\mu\nu}~
        \delta h^{\mu\nu}     \ ,
\end{eqnarray}
where
\begin{eqnarray}
&&
{\cal U}_{\mu\nu}=
	-\frac{1}{2}h_{\mu\nu}\eta^{|\alpha}\eta_{|\alpha}
	+\eta_{|\mu}\eta_{|\nu} \ ,\\
&&
{\cal Q}_{\mu\nu}=
	V\left(R_{\mu\nu}-\frac{1}{2}h_{\mu\nu}R\right)
	-(~V''~\eta_{|\mu}\eta_{|\nu}+V'~\eta_{|\mu\nu}~)
	+h_{\mu\nu}(~V''~\eta^{|\alpha}\eta_{|\alpha}+V'~\Box\eta~) \ .
\end{eqnarray}
These formula can be used to read off action from the Einstein equations.
  
Using Eq.~(\ref{conservation}), the above quantities satisfy the following
 identities
\begin{eqnarray}
{\cal S}_{\mu\nu}{}^{|\nu}&=&0 \ ,\\
{\cal L}_{\mu\nu}{}^{|\nu}&=&
	R^{|\alpha}\eta_{|\alpha}\eta_{|\mu}
	+R(\Box\eta)\eta_{|\mu} \ ,\\
{\cal J}_{\mu\nu}{}^{|\nu}&=&
	\frac{1}{2}R^{|\alpha}\eta_{|\alpha}\eta_{|\mu}
	+R^{\alpha\beta}\eta_{|\alpha\beta}\eta_{|\mu} \ ,\\
{\cal P}_{\mu\nu}{}^{|\nu}&=&
	-(\Box^2\eta)\eta_{|\mu} \ ,\\
{\cal M}_{\mu\nu}{}^{|\nu}&=&
	4~\eta_{|\alpha\beta}\eta^{|\alpha}\eta^{|\beta}\eta_{|\mu}
	+2~\eta^{|\alpha}\eta_{|\alpha}(\Box\eta)\eta_{|\mu} \ , \\
{\cal U}_{\mu\nu}{}^{|\nu}&=&
	(\Box\eta)\eta_{|\mu} \ , \\
{\cal Q}_{\mu\nu}{}^{|\nu}&=&
	-\frac{1}{2}R~V'~\eta_{|\mu} \ . 
\end{eqnarray}
These identities are useful to check the momentum constraint and
 the Bianchi identity. 

It is also useful to know the traces of these quantities
\begin{eqnarray}
&&\qquad\qquad
{\cal S}=0 \ ,\\
&&\qquad\qquad
{\cal L}=3~\Box(\eta^{|\alpha}\eta_{|\alpha}) \ ,\\
&&\qquad\qquad
{\cal J}=\frac{1}{2}\Box(\eta^{|\alpha}\eta_{|\alpha})
	+(\eta^{|\alpha}\eta^{|\beta})_{|\alpha\beta} \ ,\\
&&\qquad\qquad
{\cal P}=2(\Box\eta)^2+2(\Box\eta)^{|\alpha}\eta_{|\alpha} \ ,\\
&&\qquad\qquad
{\cal M}=0 \ ,\\
&&\qquad\qquad
{\cal U}=-\eta^{|\alpha}\eta_{|\alpha} \ ,\\
&&\qquad\qquad
{\cal Q}=-VR
	+3V''~\eta^{|\alpha}\eta_{|\alpha}
	+3V'~\Box\eta \ .
\end{eqnarray}

\section{Second order calculations}

In this Appendix we show the solutions of basic equations 
(\ref{eq:hamiltonian})-(\ref{eq:scalar}) at  second order
in our iteration scheme. 
We  take $\chi_{\mu\nu} = C=0$, which is consistent
with the results in the single-brane limit, $\Psi\rightarrow1$, 
for the 2-brane system. 
 
The Hamiltonian constraint (\ref{eq:hamiltonian}) at second order 
takes the form
\begin{eqnarray}
\overset{(2)}{K}=-{\ell\over 8}\overset{(1)}{K}{}^2
	+{\ell\over 6}\overset{(1)}{\Sigma}{}^\alpha{}_\beta
	\overset{(1)}{\Sigma}{}^\beta{}_\alpha
	+{\ell\over 6}\left[ \overset{(4)}{R}
	-\kappa^2\nabla^\alpha\varphi\nabla _\alpha\varphi\right]^{(2)}
	+{\ell\over 6}\kappa^2\left(
	\partial_y \overset{(1)}{\varphi} \right)^2 
	-{\ell\kappa^2\over 3}V^\prime\overset{(1)}{\varphi}\ .
	\label{2:hamiltonian}
\end{eqnarray}
Substituting the solutions (\ref{1:hamiltonian}) -- (\ref{1:scalar}) 
into above equation (\ref{2:hamiltonian}), we get
\begin{eqnarray}
\overset{(2)}{K} &=& \frac{\ell^3}{b^4}\left[ -\frac{1}{72}R^2
	+\frac{1}{24}R^{\alpha}{}_{\beta}R^{\beta}{}_{\alpha}
	+\frac{\kappa^2}{36}R\eta^{|\alpha}\eta_{|\alpha}
	-\frac{\kappa^2}{12}R^{\alpha\beta}\eta_{|\alpha}\eta_{|\beta}
	+\frac{\kappa^2}{24}(\Box\eta)^2
	+\frac{\kappa^4}{36}\left(\eta^{|\alpha}\eta_{|\alpha}\right)^2
	\right]
	\nonumber \\
&&	+\frac{\ell^3}{b^2}\kappa^2\left[
	\frac{1}{72}RV-\frac{\kappa^2}{72}V\eta^{|\alpha}\eta_{|\alpha}
	-\frac{1}{8}V'\Box\eta\right] \nonumber\\
&&	+\frac{\ell^3}{12}\left(\frac{1}{b^4}-\frac{1}{b^2}\right)
	\left[ R^{\alpha}{}_{\beta}R^{\beta}{}_{\alpha}
	-\frac{1}{6}R^2 
	+\kappa^2(\Box\eta)^2 
	-2\kappa^2R^{\alpha\beta}\eta_{|\alpha}\eta_{|\beta}
	+\frac{\kappa^2}{3}R\eta^{|\alpha}\eta_{|\alpha}
	+\frac{5}{6}\kappa^4
	\left(\eta^{|\alpha}\eta_{|\alpha}\right)^2\right]
	\nonumber\\
&&	-\frac{\ell^3}{72}\kappa^4V^2
	+\frac{\ell^3}{96}\kappa^2V^{\prime 2} 
	+\frac{\ell^3}{12}\kappa^2 V^\prime \Box\eta
	\nonumber\\
&&	+\frac{\ell^3}{12}\kappa^2\frac{\log b}{b^2}
	\left[ V'\Box\eta
	-\frac{\kappa^2}{3}V\eta^{|\alpha}\eta_{|\alpha}
	+\frac{1}{3}RV \right] 
	-\frac{\ell^3}{12}\kappa^2 V^{\prime 2} \log b \ .
	\label{2:trace}
\end{eqnarray}
The evolution equation (\ref{eq:evolution}) becomes
\begin{eqnarray}
\overset{(2)}{\Sigma}{}^\mu{}_{\nu,y} 
	-{4\over\ell}\overset{(2)}{\Sigma}{}^\mu{}_\nu 
	=\overset{(1)}{K} \overset{(1)}{\Sigma}{}^\mu{}_\nu 
 	-\left[\overset{(4)}{R}{}^{\mu}{}_{\nu} 
	-\kappa^2\nabla^\mu\varphi\nabla _\nu\varphi\right]_
	{\rm traceless}^{(2)} \ .
 	\label{2:evolution}
\end{eqnarray}
We can integrate Eq.~(\ref{2:evolution}) easily as
\begin{eqnarray}
\overset{(2)}{\Sigma}{}^{\mu}{}_{\nu} &=& 
	-\frac{\ell^3}{12}\frac{\log b}{b^4}
	\left(R-\kappa^2\eta^{|\alpha}\eta_{|\alpha}\right)
	\left[R^{\mu}{}_{\nu}-\frac{1}{4}\delta^{\mu}_{\nu}R
	-\kappa^2\eta^{|\mu}\eta_{|\nu}
	+\frac{\kappa^2}{4}\delta^{\mu}_{\nu}
	\eta^{|\alpha}\eta_{|\alpha}\right]
	\nonumber\\
&&	+\frac{\ell^3}{12}\frac{1}{b^2}\kappa^2V
	\left[R^{\mu}{}_{\nu}-\frac{1}{4}\delta^{\mu}_{\nu}R
	-\kappa^2\eta^{|\mu}\eta_{|\nu}
	+\frac{\kappa^2}{4}\delta^{\mu}_{\nu}
	\eta^{|\alpha}\eta_{|\alpha}\right]
	\nonumber\\
&&	-\frac{\ell^3}{4}\left(\frac{\log b}{b^4}-\frac{1}{2b^2}\right)
	\left[-2{\cal S}^{\mu}{}_{\nu}
	+2\kappa^2{\cal J}^{\mu}{}_{\nu}
	-\frac{2}{3}\kappa^2{\cal L}^{\mu}{}_{\nu}
	-\kappa^2{\cal P}^{\mu}{}_{\nu}
	-\frac{5}{6}\kappa^4{\cal M}^{\mu}{}_{\nu} 
	\right.\nonumber\\
&&\qquad\left.
	-\frac{1}{3}RR^{\mu}{}_{\nu}
	+\frac{\kappa^2}{3}R^{\mu}{}_{\nu}\eta^{|\alpha}\eta_{|\alpha}
	+\frac{\kappa^2}{3}R~\eta^{|\mu}\eta_{|\nu}
	\right]_{\rm traceless}
	\nonumber\\
&&	+\frac{\ell^3}{12}\kappa^2
	\left(\frac{\log b}{b^2}-\frac{1}{2b^2}\right)
	\left[-3V''\eta^{|\mu}\eta_{|\nu}
	-\kappa^2V\eta^{|\mu}\eta_{|\nu}
	-{\cal Q}^{\mu}{}_{\nu}
	+2VR^{\mu}{}_{\nu}
	\right]_{\rm traceless}
	+\frac{\overset{(2)}{\chi}{}^{\mu}{}_{\nu}}{b^4} \ .
	\label{2:traceless}
\end{eqnarray}
Finally,  the equation for the scalar field (\ref{eq:scalar}) gives
\begin{eqnarray}
\partial_y^2\overset{(2)}{\varphi} 
	-{4\over \ell}\partial_y \overset{(2)}{\varphi}
	=\overset{(1)}{K} \partial_y \overset{(1)}{\varphi}
 	-\left[\nabla^{\alpha}\nabla_{\alpha}\varphi\right]^{(2)}
 	+V^{\prime\prime}\overset{(1)}{\varphi} \ .
\end{eqnarray}
The result of integration becomes
\begin{eqnarray}
\partial_y\overset{(2)}{\varphi} &=& \frac{\stackrel{(2)}{C}}{b^4}
	-\frac{\ell^3}{12}\frac{\log b}{b^4}\left[
	\left( R-\kappa^2\eta^{|\alpha}\eta_{|\alpha}\right) 
	\Box\eta
	-3\Box^2\eta
	-6R^{\alpha\beta}\eta_{|\alpha\beta}
	+R\Box\eta
	-R^{|\alpha}\eta_{|\alpha}
	+8\kappa^2\eta^{|\alpha}\eta^{|\beta}\eta_{|\alpha\beta}
	+5\kappa^2\eta^{|\alpha}\eta_{|\alpha}\Box\eta\right]
	\nonumber\\
&&	+\frac{\ell^3}{48}\frac{1}{b^2}\left[ \left( R
	-\kappa^2\eta^{|\alpha}\eta_{|\alpha}\right) V'
	+4\ell^3\kappa^2V\Box\eta
	-3\ell^3(V'''-\frac{2}{3}\kappa^2V')\eta^{|\alpha}\eta_{|\alpha}
	-3\ell^3(V''+\frac{2}{3}\kappa^2V)\Box\eta
	-6\ell^3V''~\Box\eta
	\right.\nonumber\\
&&\qquad\left.
	-6\Box^2\eta
	-12R^{\alpha\beta}\eta_{|\alpha\beta}
	+2R\Box\eta
	-2R^{|\alpha}\eta_{|\alpha}
	+16\kappa^2\eta^{|\alpha}\eta^{|\beta}\eta_{|\alpha\beta}
	+10\kappa^2\eta^{|\alpha}\eta_{|\alpha}\Box\eta\right]
	\nonumber\\
&&	-\frac{\ell^3}{48}\kappa^2VV'
	+\frac{\ell^3}{16}V''\Box\eta+\frac{\ell^3}{64}V'V''
	+\frac{\ell^3}{8}\frac{\log b}{b^2}
	\left[(V'''-\frac{2}{3}\kappa^2V')\eta^{|\alpha}\eta_{|\alpha}
	+(V''+\frac{2}{3}\kappa^2V)\Box\eta \right]
	\nonumber\\
&&	-\frac{\ell^3}{16}\log b~V''V' \ .
	\label{2:scalar}
\end{eqnarray}
Similar to results at  first order, we get two kinds of constant of 
integration
$\overset{(2)}{\chi}{}^{\mu}{}_{\nu}$ and $\overset{(2)}{C}$.
Momentum constraint (\ref{eq:momentum}) at this order
\begin{eqnarray}
\overset{(2)}{\Sigma}{}^{\nu}{}_{\mu|\nu}
	-{3\over 4}\overset{(2)}{K}{}_{|\mu}
	+\overset{(1)}{\Gamma}{}^{\nu}_{\alpha\nu}
	\overset{(1)}{\Sigma}{}^{\alpha}{}_{\mu}   
	-\overset{(1)}{\Gamma}{}^{\alpha}_{\nu\mu}
	\overset{(1)}{\Sigma}{}^{\nu}{}_{\alpha} 
	=-\kappa^2\left(\partial_y \overset{(2)}{\varphi}\right)\eta_{|\mu}
	-\kappa^2\left(\partial_y\overset{(1)}{\varphi}\right)
	\overset{(1)}{\varphi}{}_{|\mu}
\end{eqnarray}
also gives the constraint on these  constants of integration
\begin{eqnarray}
&&\overset{(2)}{\chi}{}^{\nu}{}_{\mu|\nu}
	+\frac{\ell^3}{24}\left[
	-RR^{\nu}{}_{\mu}
	+\kappa^2R^{\nu}{}_{\mu}\eta^{|\alpha}\eta_{|\alpha}
	+\kappa^2R\eta^{|\nu}\eta_{|\mu}
	\right.\nonumber\\
&&\qquad\left.
	+\frac{1}{2}\delta^{\nu}_{\mu}
	\left(R^2
	-\frac{3}{2}R^{\alpha}_{\beta}R^{\beta}_{\alpha}
	-2\kappa^2R\eta^{|\alpha}\eta_{|\alpha}
	+3\kappa^2R^{\alpha\beta}\eta_{|\alpha}\eta_{|\beta}
	-\frac{3\kappa^2}{2}(\Box\eta)^2
	-\kappa^4(\eta^{|\alpha}\eta_{|\alpha})^2\right) 
	\right]_{|\nu}
	\nonumber\\
&&\qquad
	=\kappa^2\left(\frac{\ell^3}{24}R\Box\eta
	+\frac{\ell^3}{12}\kappa^2
	\eta^{|\alpha}\eta^{|\beta}\eta_{|\alpha\beta}
	-\overset{(2)}{C}\right)\eta_{|\mu} \ .
	\label{2:mc}
\end{eqnarray}
Substituting the solutions (\ref{2:trace}), (\ref{2:traceless}) and 
(\ref{2:scalar}) into the junction conditions, then
the junction condition up to  second order gives us the effective 
equations with KK corrections. 

As we solved the bulk equations, 
 we can calculate $E_{\mu\nu}$  and $\Phi_2$ explicitly.  
The projected Weyl tensor is calculated  by
\begin{eqnarray}
E^{\mu}{}_{\nu} &=& \left[
	\left( \Sigma^{\mu}{}_{\nu}
	-{3\over 4}\delta^\mu_{\nu} K\right)_{,y}
	-{1\over 2} K \Sigma^{\mu}{}_{\nu}
	+{3\over 16}\delta^\mu_\nu K^2 
	- \Sigma^{\mu}{}_{\alpha}\Sigma^{\alpha}{}_{\nu}
	+ \delta^{\mu}_{\nu}\Sigma^{\alpha}{}_{\beta}\Sigma^{\beta}{}_{\alpha}
	\right.\nonumber\\
	&& \quad
	\left.
	-{\kappa^2 \over 3} \partial^\mu \varphi \partial_\nu \varphi
   	+{\kappa^2 \over 12}\delta^{\mu}_{\nu} 
   	\partial^\alpha \varphi \partial_\alpha \varphi
	+{3\kappa^2\over 4}\delta^{\mu}_{\nu}\left(\partial_y\varphi\right)^2
   	+{\kappa^2 \over 2}\delta^{\mu}_{\nu}V (\varphi ) 
   	\right]\Bigg|_{y=0} 
   	\label{weyl}
\end{eqnarray}
and the correspondent quantity for the scalar field is defined by
\begin{eqnarray}
\Phi_2 = \left[
	\partial^2_y \varphi
	\right] \Big|_{y=0}  \ .
\end{eqnarray}
Using these formulas, we get the above quantities at second order
\begin{eqnarray}
\overset{(2)}{E}{}^{\mu}{}_{\nu} &=& 
	\frac{\ell^2}{4}\left[
	-2{\cal S}^{\mu}{}_{\nu}
	+2\kappa^2{\cal J}^{\mu}{}_{\nu}
	-\frac{2}{3}\kappa^2{\cal L}^{\mu}{}_{\nu}
	-\kappa^2{\cal P}^{\mu}{}_{\nu}
	-\frac{5}{6}\kappa^4{\cal M}^{\mu}{}_{\nu}
	\right]
	\nonumber\\
&&	+\frac{\ell^2}{12}\kappa^2
	\left[{\cal Q}^{\mu}{}_{\nu}
	+3V''{\cal U}^{\mu}{}_{\nu}
	-\frac{3}{4}\delta^{\mu}_{\nu}V'\Box\eta
	-V(R^{\mu}{}_{\nu}-\frac{1}{2}\delta^{\mu}_{\nu}R)
	\right]
	\nonumber\\
&&	-\frac{\ell^2}{12}\kappa^4\eta^{|\alpha}\eta_{|\alpha}
	\left(
	\eta^{|\mu}\eta_{|\nu}
	-\frac{1}{4}\delta^{\mu}_{\nu}\eta^{|\alpha}\eta_{|\alpha}
	\right)
	+\frac{\ell^2}{4}\left(
	\frac{1}{3}RR^{\mu}{}_{\nu}
	-\frac{1}{12}\delta^{\mu}_{\nu}R^2
	-R^{\mu\alpha}R_{\alpha\nu}
	+\frac{1}{4}\delta^{\mu}_{\nu}R^{\alpha}{}_{\beta}R^{\beta}{}_{\alpha}
	\right)
	\nonumber\\
&&	-\frac{\ell^2}{12}\kappa^2\left(
	R\eta^{|\mu}\eta_{|\nu}+R^{\mu}{}_{\nu}\eta^{|\alpha}\eta_{|\alpha}
	\right)
	+\frac{\ell^2}{4}\kappa^2\left(
	R^{\mu\alpha}\eta_{|\alpha}\eta_{|\nu}
	+R_{\alpha\nu}\eta^{|\mu}\eta^{|\alpha}
	\right)
	\nonumber\\
&&	+\frac{\ell^2}{24}\kappa^2\delta^{\mu}_{\nu}
	R\eta^{|\alpha}\eta_{|\alpha}
	-\frac{\ell^2}{8}\kappa^2\delta^{\mu}_{\nu}
	R^{\alpha\beta}\eta_{|\alpha}\eta_{|\beta}
	+\frac{2}{\ell}\overset{(2)}{\chi}{}^{\mu}{}_{\nu}
	\label{2:weyl}
\end{eqnarray}
and
\begin{eqnarray}
\overset{(2)}{\Phi}_2&=&
	+\frac{\ell^2}{4}\left[
	\left(R+3\kappa^2\eta^{|\alpha}\eta_{|\alpha}\right)\Box\eta
	-2\Box^2\eta
	-4R^{\alpha\beta}\eta_{|\alpha\beta}
	-\frac{2}{3}R^{|\alpha}\eta_{|\alpha}
	+\frac{16}{3}\kappa^2\eta^{|\alpha}\eta^{|\beta}\eta_{|\alpha\beta}
	\right] \nonumber\\
&&
	+\frac{\ell^2}{24}RV'
	-\frac{\ell^2}{4}(V'''
	-\frac{\kappa^2}{2}V')\eta^{|\alpha}\eta_{|\alpha}
	-\frac{\ell^2}{2}V''\Box\eta
	+\frac{\ell^2}{16}V'V''
	+\frac{4}{\ell}\overset{(2)}{C} \ .
	\label{2:phi2}
\end{eqnarray}
Using these result Eqs.~(\ref{2:weyl}) and (\ref{2:phi2}), we can discuss
the relation between AdS/CFT approach and the geometrical approach.


\end{document}